\documentclass[twoside]{article}
\usepackage[a4paper]{geometry}
\usepackage[latin1]{inputenc} % ou \usepackage[utf8]{inputenc}
\usepackage[T1]{fontenc} % ou \usepackage[OT1]{fontenc}
\usepackage{RR}
\usepackage{hyperref}
\usepackage{graphicx}
\usepackage{epsfig}
\usepackage{subfigure}
\usepackage{amsmath,amsgen,amstext,amssymb}
\usepackage{amsfonts}
\usepackage{latexsym}

\newcommand{\beq}{\begin{equation}}
\newcommand{\deq}{\end{equation}}

\newcommand{\beqm}{\begin{equation*}}
\newcommand{\deqm}{\end{equation*}}

\newcommand{\baq}{\begin{eqnarray}}
\newcommand{\daq}{\end{eqnarray}}

\newcommand{\baqm}{\begin{eqnarray*}}
\newcommand{\daqm}{\end{eqnarray*}}

\newcommand{\qed}{\hfill \mbox{\raggedright $\square$}}

\newcommand{\ones}{\underline{1}}

\newcommand{\eps}{\varepsilon}

\newtheorem{thm}{Theorem}

\newtheorem{algo}{Algorithm}

\RRdate{February 2012}
\RRauthor{% les auteurs
Konstantin Avrachenkov\thanks{INRIA Sophia Antipolis, France, K.Avrachenkov@sophia.inria.fr}%
\and
Nelly Litvak\thanks{University of Twente, the Netherlands, N.Litvak@utwente.nl}%
\and 
Marina Sokol\thanks{INRIA Sophia Antipolis, France, Marina.Sokol@sophia.inria.fr}
\and\\
Don Towsley\thanks{University of Massachusetts Amherst, USA, towsley@cs.umass.edu}
}
\authorhead{K. Avrachenkov \& N. Litvak \& M. Sokol \& D. Towsley}
\RRtitle{D\'etection Rapide de Noeuds \`a Degr\'es \'Elev\'es}
\RRetitle{Quick Detection of Nodes with Large Degrees}
\titlehead{Quick Detection of Nodes with Large Degrees}
\RRresume{
Notre objectif est de trouver rapidement dans les grands r\'eseaux complexes 
top $k$ listes de noeuds avec les plus grands degr\'es. Si la liste d'adjacence du r\'eseau est connu 
(pas souvent le cas dans les r\'eseaux complexes), un algorithme d\'eterministe pour trouver un noeud 
avec le plus grand degr\'e n\'ecessite une complexit\'e moyenne de $\mbox{O}(n)$, o\`u $n$ est le nombre 
de noeuds dans le r\'eseau. M\^eme cette complexit\'e modeste peut \^etre tr\`es \'elev\'e pour les grands 
r\'eseaux complexes. Nous proposons d'utiliser une m\'ethode bas\'e sur le marche al\'eatoire. 
Nous montrons th\'eoriquement et par exp\'erimentations num\'eriques que pour les grands r\'eseaux 
la m\'ethode de marche al\'eatoire trouve top $k$ listes de bonne qualit\'e avec une forte probabilit\'e
de r\'eussite et avec des \'economies de calcul de plusieurs ordres de grandeur. Nous proposons 
\'egalement des crit\`eres d'arr\^et pour la m\'ethode de marche al\'eatoire qui ne n\'ecessite pas
de connaissance de la structure du r\'eseau.
}
\RRabstract{
Our goal is to quickly find top $k$ lists of nodes with the largest degrees in large complex networks.
If the adjacency list of the network is known (not often the case in complex networks),
a deterministic algorithm to find a node with the largest degree requires an
average complexity of $\mbox{O}(n)$, where $n$ is the number of nodes in the network.
Even this modest complexity can be very high for large complex networks.
We propose to use the random walk based method. We show theoretically and by
numerical experiments that for large networks the random walk method finds
good quality top lists of nodes with high probability and with computational
savings of orders of magnitude. We also propose stopping criteria for the random
walk method which requires very little knowledge about the structure of the network.
}
\RRmotcle{r\'eseaux complexes, d\'etection de noeuds avec les plus grands degr\'es,  
top $k$ liste, marche al\'eatoire, crit\`eres d'arr\^et}
\RRkeyword{Complex networks, detection of nodes with the largest degrees, top $k$ list,
random walk, stopping criteria}
\RRprojet{Maestro}  % cas d'un seul projet
\RCSophia % Sophia Antipolis M\'editerran\'ee

\begin{document}
\RRNo{7881}
\makeRR   % cas d'un rapport de recherche

\section{Introduction}
\label{sec:intro}
We are interested in quickly detecting nodes with large degrees in very large networks.
Firstly, node degree is one of centrality measures used for the analysis of complex
networks. Secondly, large degree nodes can serve as proxies for central nodes
corresponding to the other centrality measures as betweenness centrality or
closeness centrality \cite{Letal11,MBW10}. In the
present work we restrict ourself to undirected networks or symmetrized versions
of directed networks. In particular, this
assumption is well justified in social networks. Typically, friendship or
acquaintance is a symmetric relation. If the adjacency list of the network
is known (not often the case in complex networks), the straightforward method that
comes to mind is to use one of the standard sorting algorithms like Quicksort
or Heapsort. However, even their modest average complexity, $\mbox{O}(n\log(n))$,
can be very high for very large complex networks.
In the present work we suggest using random walk based methods for detecting
a small number of nodes with the largest degree. The main idea is that
the random walk very quickly comes across large degree nodes. In our numerical
experiments random walks outperform the standard sorting procedures by orders
of magnitude in terms of computational complexity. For instance, in our experiments
with the web graph of the UK domain (about 18\,500\,000 nodes) the random walk
method spends on average only about 5\,400 steps to detect the largest degree node.
Potential memory savings are also significant since the method does not require
knowledge of the entire network.
In many practical applications we do not need a complete ordering of the nodes
and even can tolerate some errors in the top list of nodes. We observe that
the random walk method obtains many nodes in the top list correctly
and even those nodes that are erroneously placed in the top list have large degrees.
Therefore, as typically happens in randomized algorithms \cite{M06,MR95},
we trade off exact results for very good approximate results or for exact results
with high probability and gain significantly in computational efficiency.

The paper is organized as follows: in the next section we introduce our basic
random walk with uniform jumps and demonstrate that it is able to quickly find large
degree nodes. Then, in Section~3 using configuration model we provide an estimate
for the necessary number of steps for the random walk. In Section~4 we propose
stopping criteria that use very little information about the network. In Section~5
we show the benefits of allowing few erroneous elements in the top $k$ list.
Finally, we conclude the paper in Section~6.

\section{Random walk with uniform jumps}
\label{sec:method}

Let us consider a random walk with uniform jumps which serves as a basic
algorithm for quick detection of large degree nodes. The random walk with uniform jumps
is described by the following transition probabilities \cite{ART10}
\begin{equation}\label{eq:probrestart}
p_{ij} = \left\{ \begin{array}{ll}
\frac{\alpha/n+1}{d_i+\alpha}, & \mbox{if $i$ has a link to $j$},\\
\frac{\alpha/n}{d_i+\alpha}, & \mbox{if $i$ does not have a link to $j$},
\end{array}\right.
\end{equation}
where $d_i$ is the degree of node $i$.
The random walk with uniform jumps can be regarded as a random walk
on a modified graph where all the nodes in the graph are connected by
artificial edges with a weight $\alpha/n$. The parameter $\alpha$ controls
the rate of jumps. Introduction of jumps helps in a number of ways.
As was shown in \cite{ART10}, it reduces the mixing time to
stationarity. It also solves a problem encountered by a random walk on a graph
consisting of two or more components, namely the inability to visit all nodes.
The random walk with jumps
also reduces the variance of the network function estimator \cite{ART10}.
This random walk resembles the PageRank random walk. However, unlike
the PageRank random walk, the introduced random walk is reversible.
One important consequence of the reversibility of the random walk is that
its stationary distribution is given by a simple formula
\begin{equation}\label{eq:modstdistr}
\pi_i(\alpha) = \frac{d_i+\alpha}{2|E|+n\alpha} \,\,\, \quad
\forall i \in V,
\end{equation}
from which the stationary distribution of the original random walk
can easily be retrieved. We observe that the modification preserves
the order of the nodes' degrees, which is particularly important for
our application.

We illustrate on several network examples how the random walk helps us
quickly detect large degree nodes. We consider as examples
one synthetic network generated by the preferential attachment rule and two
natural large networks. The Preferential Attachment (PA) network combines 100\,000
nodes. It has been generated according to the generalized preferential attachment
mechanism \cite{DMS00}. The average degree
of the PA network is two and the power law exponent is 2.5.
The first natural example is the symmetrized web
graph of the whole UK domain crawled in 2002 \cite{BV04}. The UK network has
18\,520\,486 nodes and its average degree is 28.6. The second natural
example is the network of co-authorships of DBLP \cite{PBetal11}. Each node represents
an author and each link represents a co-authorship of at least one article.
The DBLP network has 986\,324 nodes and its average degree is 6.8.

We carry out the following experiment: we initialize the random walk (\ref{eq:probrestart})
at a node chosen according to the uniform distribution and continue the random walk
until we hit the largest degree node. The largest degrees for the PA, UK and
DBLP networks are 138, 194\,955, and 979, respectively. For the PA network we have made
10\,000 experiments and for the UK and DBLP networks we performed 1\,000 experiments
(these networks were too large to perform more experiments).

\begin{figure}[ht]
  \begin{center}
  \subfigure[$\alpha = 0$]{
    \includegraphics[scale=0.37]{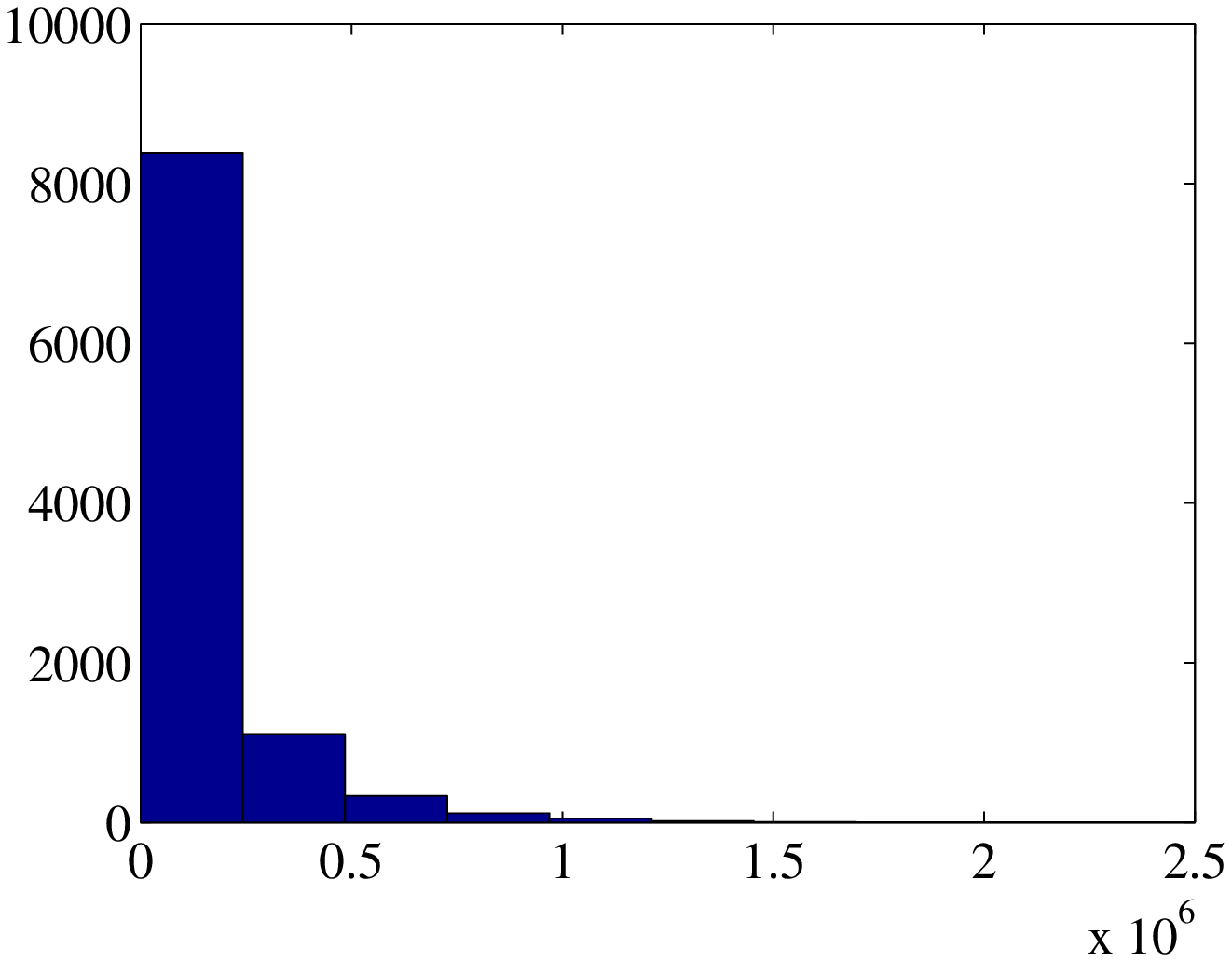}
  }
   \subfigure[$\alpha = 2$]{
   \includegraphics[scale=0.37]{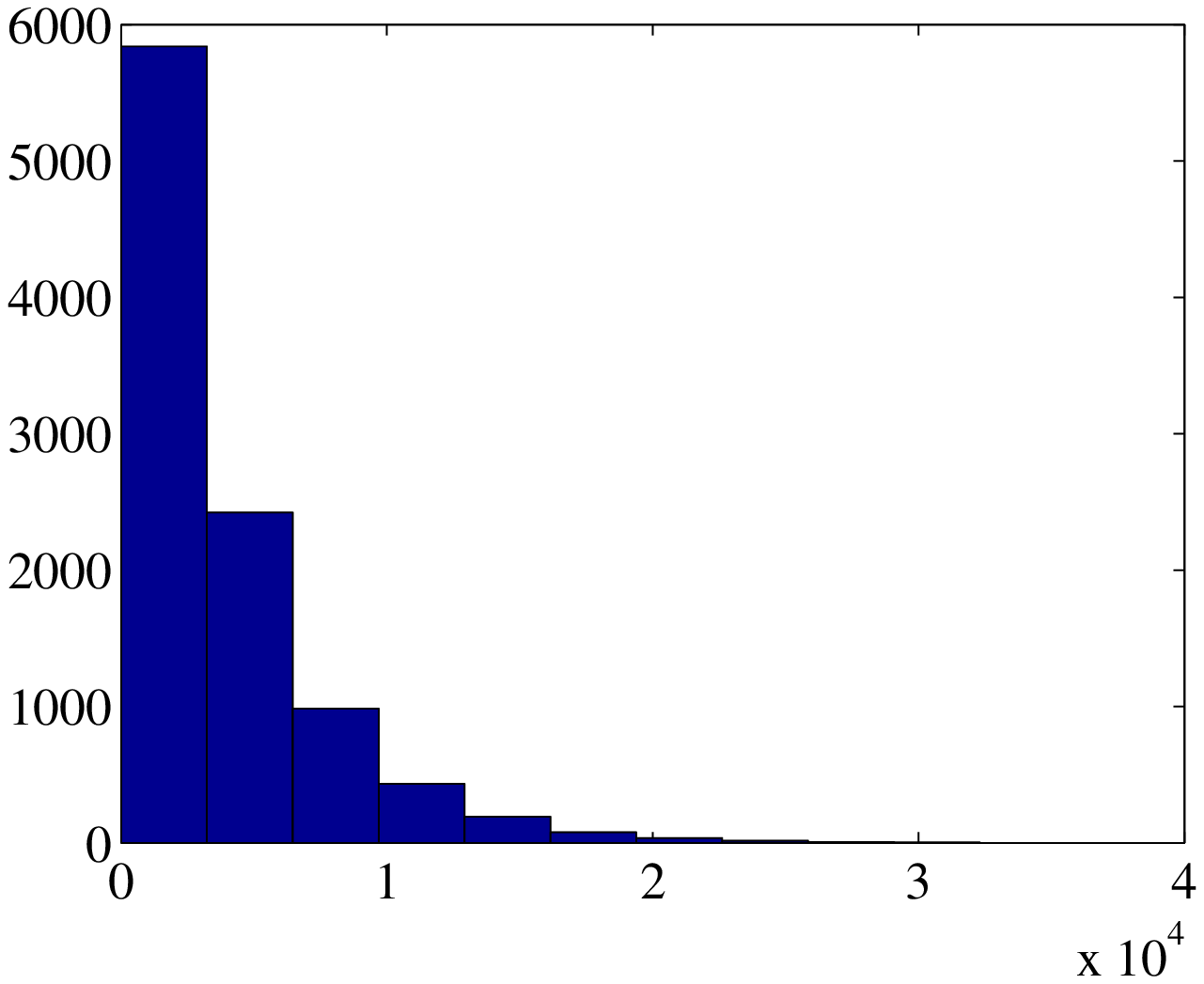}
   }
  \caption{Histograms of hitting times in the PA network.\label{fig:HistHitTimePA}}
  \end{center}
\end{figure}

In Figue~\ref{fig:HistHitTimePA} we plot the histograms of hitting times for the PA network.
The first remarkable observation is that when $\alpha=0$ (no restart) the average hitting time,
which is equal to 123\,000, is nearly three orders of magnitude larger than 3\,720,
the hitting time when $\alpha=2$. The second remarkable observation is that 3\,720 is not
too far from the value
$$
1/\pi_{max}(\alpha) = (2|E|+n\alpha)/(d_{max}+\alpha) = 2\,857,
$$
which corresponds to the average return time to the largest degree node in the random walk
with jumps.

We were not able to collect a representative number of experiments for the UK and DBLP networks
when $\alpha=0$. The reason for this is that the random walk gets stuck either in disconected or
weakly connected components of the networks. For the UK network we were able to make 1\,000
experiments with $\alpha=0.001$ and obtain the average hitting time 30\,750. Whereas if we
take $\alpha=28.6$ for the UK network, we obtain the average hitting time 5\,800. Note that
the expected return time to the largest degree node in the UK network is given by
$$
1/\pi_{max}(\alpha) = (2|E|+n\alpha)/(d_{max}+\alpha) = 5\,432 .
$$
For the DBLP graph we conducted 1\,000 experiments with $\alpha=0.00001$ and obtained
an average hitting time of 41\,131. Whereas if we take $\alpha=6.8$, we obtain an average hitting
time of 14\,200. The expected return time to the largest degree node in the DBLP network is given by
$$
1/\pi_{max}(\alpha) = (2|E|+n\alpha)/(d_{max}+\alpha) = 13\,607 .
$$
The two natural network examples confirm our guess that the average hitting time for the largest
degree node is fairly close to the average return time to the largest degree node. Let us also
confirm our guess with asymptotic analysis.

\begin{thm}
Without loss of generality, index the nodes such that node 1 has the largest degree,
$(1,i) \in E, i=2,...,s, s=d_1+1$, and let $\nu$ denote the initial distribution of the random walk with jumps.
Then, the expected hitting time to node 1 starting from any initial distribution $\nu$ is given by
\begin{equation} \label{eq:hittimeasymp}
E_\nu[T_1] = \frac{\sum_{i=2}^n d_i + (n - 1)\alpha}{d_1+2\alpha(1-1/n)}
+ \mbox{o}\left(\min_{i=2,...,s}\{(d_i+\alpha),n\}\right),
\end{equation}
\end{thm}
{\bf Proof:}
The expected hitting time from distribution $\nu$ to node 1 is given by the formula
\begin{equation}\label{eq:hittime1}
E_\nu[T_1] = \nu [I-P_{-1}]^{-1} \ones,
\end{equation}
where $P_{-1}$ is a taboo probability matrix (i.e., matrix $P$ with the $1$-st row
and $1$-st column removed). The matrix $P_{-1}$ is substochastic but is
very close to stochastic. Let us represent it as a stochastic matrix minus some perturbation term:
$$
P_{-1} = \tilde{P}-\eps Q = \tilde{P}-
\left[
  \begin{array}{cccccc}
    \frac{1+2\alpha/n}{d_2+\alpha} & 0 &  &  &  & 0 \\
    0 & \ddots &  &  &  &  \\
     &  & \frac{1+2\alpha/n}{d_s+\alpha} &  &  &  \\
     &  &  & \frac{2\alpha/n}{d_{s+1}+\alpha} &  &  \\
     &  &  &  & \ddots & 0 \\
    0 &  &  &  & 0 & \frac{2\alpha/n}{d_n+\alpha} \\
  \end{array}
\right]
$$
We add missing probability mass to the diagonal of $\tilde{P}$, which corresponds to an increase
in the weights for self-loops. The matrix $\tilde{P}$ represents a reversible Markov chain with the stationary
distribution
$$
\tilde \pi_j = \frac{d_j+\alpha}{\sum_{i=2}^n d_i+(n-1)\alpha}.
$$
Now we can use the following result from the perturbation theory (see Lemma~1 in \cite{ABN10}):
\begin{equation}\label{eq:laurent}
[I-\tilde{P}+\eps Q]^{-1} = \frac{\ones \tilde\pi}{\tilde\pi(\eps Q)\ones}+X_0+\eps X_1+... \ ,
\end{equation}
where $\tilde\pi$ is the stationary distribution of the stochastic matrix $\tilde{P}$.
In our case, the quantity $\max_{i=2,...,s}\{1/(d_i+\alpha),1/n\}$ will play the role of $\eps$.
We apply the series (\ref{eq:laurent}) to approximate the expected hitting time.
Towards this goal, we calculate
$$
\tilde\pi (\eps Q) \ones = \sum_{j=2}^n \tilde\pi_j \eps q_{jj}
$$
$$
= \sum_{j=2}^s \frac{d_j+\alpha}{\sum_{i=2}^n d_i+(n-1)\alpha} \frac{1+2\alpha/n}{d_j+\alpha}
+ \sum_{j=s+1}^n \frac{d_j+\alpha}{\sum_{i=2}^n d_i+(n-1)\alpha} \frac{2\alpha/n}{d_j+\alpha}
$$
$$
=\frac{d_1(1+2\alpha/n)+(n-d_1-1)(2\alpha/n)}{\sum_{i=2}^n d_i+(n-1)\alpha}
=\frac{d_1+2\alpha(1-1/n)}{\sum_{i=2}^n d_i+(n-1)\alpha}.
$$
Observing that $\nu \ones \tilde\pi \ones=1$, we obtain (\ref{eq:hittimeasymp}).

\qed

Indeed, the asymptotic expression (\ref{eq:hittimeasymp}) is very close to
$(2|E|+n\alpha)/(d_1+\alpha)$, which is the expected return time to node 1.

Based on the notion of the hitting time we propose an efficient method
for quick detection of the top $k$ list of largest degree nodes.
The algorithm maintains a top $k$ candidate list. Note that once
one of the $k$ nodes with the largest degrees appears in this candidate
list, it remains there subsequently. Thus, we are interested in hitting events.
We propose the following algorithm for detecting the top $k$ list of largest degree nodes.

\begin{algo} {\bf Random walk with jumps and candidate list}
\begin{enumerate}
\item Set $k$, $\alpha$ and $m$.
\item Execute a random walk step according to $(\ref{eq:probrestart})$.
\item Check if the current node has a larger degree than one of the nodes
in the current top $k$ candidate list. If it is the case,
insert the new node in the top-k candidate list and remove the worst node out
of the list.
\item If the number of random walk steps is less than $m$, return to
Step~2 of the algorithm. Stop, otherwise.
\end{enumerate}
\end{algo}

The value of parameter $\alpha$ is not crucial. In our experiments, we have
observed that as long as the value of $\alpha$ is neither too small nor not too
big, the algorithm performs well. A good option for the choice
of $\alpha$ is a value slightly smaller than the average node degree.
Let us explain this choice by calculating a probability of jump in the steady state
$$
\sum_{j=1}^n \pi_j(\alpha) \frac{\alpha}{d_j+\alpha}
= \sum_{j=1}^n \frac{d_j+\alpha}{2|E|+n\alpha} \frac{\alpha}{d_j+\alpha}
= \frac{n\alpha}{2|E|+n\alpha} = \frac{\alpha}{2|E|/n+\alpha}.
$$
If $\alpha$ is equal to $2|E|/n$, the average degree, the random walk
will jump in the steady state on average every two steps. Thus, if we set
$\alpha$ to the average degree or to a slightly smaller value, on one hand
the random walk will quickly converge to the steady state and on the other hand
we will not sample too much from the uniform distribution.

The number of random walk steps, $m$, is a crucial parameter.
Our experiments indicate that we obtain a top $k$ list with many correct elements
with high probability if we take the number of random walk steps to be
twice or thrice as large as the expected hitting time of the nodes
in the top $k$ list.
From Theorem~1 we know that the hitting time of the large degree node
is related to the value of the node's degree.
Thus, the problem of choosing $m$ reduces to the problem of estimating
the values of the largest degrees. We address this problem in the following
section.

\section{Estimating the largest degrees in the configuration network model}

The estimations for the values of the largest degrees can be derived
in the configuration network model \cite{HofstadRG} with a power law degree distribution.
In some applications the knowledge of the power law parameters might be available to us.
For instance, it is known that web graphs have power law degree distribution
and we know typical ranges for the power law parameters.

We assume that the node degrees $D_1,\ldots,D_n$ are i.i.d. random variables
with a power law distribution $F$ and finite expectation $E[D]$.
Let us determine the number of links contained in the top $k$ nodes. Denote
\[F(x)=P[D\le x],\quad \bar F(x)=1-F(x),\quad x\ge 0.\]
Further let $D_{(1)}\ge\ldots\ge D_{(n)}$ be the order statistics of $D_1,\ldots,D_n$. Under the assumption that $D_j$'s obey a power law, we use the results from the extreme value theory as presented in  \cite{Matthys2003quantile}, to state that there exist sequences of constants $(a_n)$ and $(b_n)$ and a constant $\delta$ such that
\begin{equation}
\label{eq:anbn}
\lim_{n\to\infty}n\bar F(a_nx+b_n)=(1+\delta x)^{-1/\delta}.
\end{equation}
This implies the following approximation for high quantiles of $F$, with exceedance probability close to zero~\cite{Matthys2003quantile}:
\[x_p\approx a_n\,\frac{(pn)^{-\delta}-1}{\delta}+b_n.\]
For the $j$th largest degree, where $j=2,\ldots,k$, the estimated exceedance probability equals $(j-1)/n$, and thus we can use the quantile $x_{(j-1)/n}$ to approximate the degree $D_{(j)}$ of this node:
\begin{equation}
\label{eq:n_links_general}
D_{(j)}\approx a_n\,\frac{(j-1)^{-\delta}-1}{\delta}+b_n.
\end{equation}

The sequences $(a_n)$ and $(b_n)$ are easy to find for a given shape of the tail of $F$. Below we derive the corresponding results for the commonly accepted Pareto tail distribution of $D$, that is,
\begin{equation}\label{eq:pareto}\bar F(t)=C x^{-\gamma}\quad \mbox{for $x>x'$},\end{equation}
where $\gamma>1$ and $x'$ is a fixed sufficiently large number so that the power law degree distribution is observed for nodes with degree larger than $x'$. In that case we have
\[\lim_{n\to\infty}n\bar F(a_nx+b_n)=\lim_{n\to\infty}n C (a_nx+b_n)^{-\gamma}=
\lim_{n\to\infty}(C^{-1/\gamma}n^{-1/\gamma}a_nx+C^{-1/\gamma}n^{-1/\gamma}b_n)^{-\gamma},\]
which directly gives (\ref{eq:anbn}) with
\begin{equation}
\label{eq:scaling}
\delta=1/\gamma,\quad a_n=\delta C^{\delta} n^{\delta},\quad b_n=C^{\delta} n^{\delta}.
\end{equation}
Substituting (\ref{eq:scaling}) into (\ref{eq:n_links_general}) we obtain the following prediction for $D_{(j)}$, $j=2,\ldots,k$, in the case of the Pareto tail of the degree distribution:
\begin{equation}
\label{eq:n_links_pareto}
D_{(j)}\approx n^{1/\gamma}[C^{1/\gamma} (j-1)^{-1/\gamma}-C^{1/\gamma} +1].
\end{equation}

It remains to find an approximation for $D_{(1)}$, the maximal degree in the graph. From the extreme value theory it is well known that if $D_1,\ldots,D_n$ obey a power law then
\[\lim_{n\to\infty}P\left(\frac{D_{(1)}-b_n}{a_n}\le x\right)=H_\delta(x)=\exp(-(1+\delta x)^{-1/\delta}),\]
where, for Pareto tail, $a_n,b_n$ and $\delta$ are defined in (\ref{eq:scaling}). Thus, as an approximation for the maximal node degree
we can choose $a_nx+b_n$ where $x$ can be chosen as either an expectation, a median or a mode of $H_\delta(x)$. If we choose the mode, $((1+\delta)^{-\delta}-1)/{\delta}$, then we obtain an approximation, which is smaller than the one for the 2nd largest degree. Further, the expectation $(\Gamma(1-\delta)-1)/\delta$ is very sensitive to the value of $\delta=1/\gamma$, especially when $\gamma$ is close to one, which is often the case in complex networks. Besides, the parameter $\gamma$ is hard to estimate with high precision. Thus, we choose the median $(\log(2))^{-\delta}-1)/\delta$, which yields
\begin{equation}
\label{eq:max}
D_{(1)}\approx a_n\,\frac{(\log(2))^{-\delta}-1}{\delta}+b_n=n^{1/\gamma}[C^{1/\gamma} (\log(2))^{-1/\gamma}-C^{1/\gamma} +1].
\end{equation}

For instance, in the PA network $\gamma=2.5$ and $C=3.7$, which gives according to (\ref{eq:max}) $D_{(1)} \approx 127$.
(This is a good prediction even though the PA network is not generated according to the configuration model.
We also note that even though the extremum distribution in the preferential attachment model is different from
that of the configuration model their ranges seem to be very close \cite{MAA02}.)
This in turn suggests that for the PA network $m$ should be chosen in the range 6\,000-18\,000 if $\alpha=2$.
As we can see from Figure~\ref{fig:NumElPA}
this is indeed a good range for the number of random walk steps. In the UK network $\gamma=1.7$ and $C=90$, which gives
$D_{(1)} \approx 82\,805$ and suggests a range of 20\,000-30\,000 for $m$ if $\alpha=28.6$.
Figure~\ref{fig:NumElUK} confirms that this is a good choice.
The degree distribution of the DBLP network does not follow a power law so we cannot apply the above reasoning to it.

%Another type of Monte Carlo algorithms that we consider are based on
%unbalanced random walk. Recall that in the balanced random walk
%(see e.g., \cite{BDX04}) the transition probabilities are roughly
%inversely proportional to the degrees of the neighbor nodes. This
%change of transition probabilities allows fast mixing to the uniform
%stationary distribution. Here, on opposite, we propose to make
%transition probabilities proportional to the neighbor nodes' degrees
%or proportional to some function of the neighbor nodes' degrees.
%Of course, the uniform restart can also be added to this method.
%
%\begin{rem}
%Isn't the latter type of the Monte Carlo algorithms just a version
%of the Metropolis-Hastings algorithm with the chosen stationary
%probability distribution $\bar\pi_i \sim d_i^2$?
%\end{rem}

\section{Stopping criteria}

Suppose now that we do not have any information about the range
for the largest $k$ degrees. In this section we design stopping criteria
that do not require knowledge about the structure of the network.
As we shall see, knowledge of the order of magnitude of the average
degree might help, but this knowledge is not imperative for a practical
implementation of the algorithm.

Let us now assume that node $j$ can be sampled independently with probability
$\pi_j(\alpha)$ as in (\ref{eq:modstdistr}). There are at least two ways to
achieve this practically. The first approach is to run the random walk for a
significant number of steps until it reaches the stationary distribution.
If one chooses $\alpha$ reasonably large, say the same order of magnitude
as the average degree, then the mixing time becomes quite small \cite{ART10}
and we can be sure to reach the stationary distribution in a small
number of steps. Then, the last step of a run of the random walk will
produce an i.i.d. sample from a distribution very close to (\ref{eq:modstdistr}).
The second approach is to run the random walk uninterruptedly, also with
a significant value of $\alpha$, and then perform Bernoulli sampling with
probability $q$ after a small initial transient phase. If $q$
is not too large, we shall have nearly independent samples following
the stationary distribution (\ref{eq:modstdistr}). In our experiment,
$q \in [0.2,0.5]$ gives good results when $\alpha$ has the same order
of magnitude as the average degree.

We now estimate the probability of detecting correctly the top $k$ list
of nodes after $m$ i.i.d. samples from (\ref{eq:modstdistr}). Denote by $X_i$
the number of hits at node $i$ after $m$ i.i.d. samples. We note that if we use
the second approach to generate i.i.d. samples, we spend approximately $m/q$
steps of the random walk. We correctly detect the top $k$ list with the
probability given by the multinomial distribution
$$
P[X_1 \ge 1,...,X_k \ge 1]=
$$
$$
\sum_{i_1 \ge 1,...,i_1 \ge 1} \frac{m!}{i_1!\cdots i_k!(m-i_1-...-i_k)!}
\pi_1^{i_1}\cdots \pi_k^{i_k} (1-\sum_{i=1}^k \pi_i)^{m-i_1-...-i_k}
$$
but it is not feasible for any realistic computations. Therefore, we propose
to use the Poisson approximation. Let $Y_j$, $j=1,...,n$ be independent
Poisson random variables with means $\pi_j m$. That is, the random variable
$Y_j$ has the following probability mass function
$P[Y_j=r]=e^{-m\pi_j} (m\pi_j)^r/r!$.
It is convenient to work with the complementary event of not detecting correctly
the top $k$ list. Then, we have
$$
P[\{X_1=0\}\cup...\cup\{X_k=0\}] \le 2 P[\{Y_1=0\}\cup...\cup\{Y_k=0\}]
$$
$$
=2(1-P[\{Y_1 \ge 1\}\cap...\cap\{Y_k \ge 1\}])=2(1-\prod_{j=1}^k P[\{Y_j \ge 1\}])
$$
\begin{equation} \label{eq:errprobtopk}
=2(1-\prod_{j=1}^k (1-P[\{Y_j = 0\}]))=2(1-\prod_{j=1}^k (1-e^{-m\pi_j}))=:a,
\end{equation}
where the first inequality follows from \cite[Thm~5.10]{M06}.
In fact, in our numerical experiments we observed that the factor 2
in the first inequality is very conservative.
For large values of $m$, the Poisson bound works very well as proper approximation.

For example, if we would like to obtain the top 10 list with at most 10\% probability
of error, we need to have on average 4.5 hits per each top element.
This can be used to design the stopping criteria for our random walk algorithm.
Let $\bar{a}\in(0,1)$ be the admissible probability of an error in the top $k$ list.
Now the idea is to stop the algorithm after $m$ steps  when the estimated value of $a$
for the first time is lower than the critical number $\bar{a}$. Clearly,
\[\hat{a}_m=2(1-\prod_{j=1}^k (1-e^{-X_j}))\]
is the maximum likelihood estimator for $a$, so we would like to choose $m$ such that $\hat{a}_m\le \bar{a}$.
The problem, however, is that we do not know which $X_j$'s are the realisations of the number of visits
to the top $k$ nodes. Then let $X_{j_1},...,X_{j_k}$ be the number of hits to the
current elements in the top k candidate list and consider the estimator
\[\hat{a}_{m,0}=2(1-\prod_{i=1}^k (1-e^{-X_{j_i}})),\]
which is the maximum likelihood estimator of the quantity
\[2(1-\prod_{i=1}^k (1-e^{-m\pi_{j_i}}))\ge a.\]
(Here $\pi_{j_i}$ is a stationary probability of the node with the score $X_{j_i}$, $i=1,\ldots,k$).
The estimator $\hat{a}_{m,0}$ is computed without knowledge of the top $k$ nodes or their degrees,
and it is an estimator of an upper bound of the estimated probability that there are errors in the top $k$ list.
This leads to the following stopping rule.

\noindent
{\it {\bf Stopping rule 0.} Stop at $m=m_0$, where
\[m_0=\arg\min\{m: \hat{a}_{m,0}\le \bar{a}\}.\]
}
\smallskip

The above stopping criterion can be simplified even further to avoid computation of $\hat{a}_{m,0}$. Since
\[\hat{a}_{m,1}:=2(1-(1-e^{-X_{j_k}})^k)\ge \hat{a}_{m,0}\ge \hat{a},\]
where $X_{j_k}$ is the number of hits of the worst element in the candidate list.
The inequality  $\hat{a}_m\le \bar{a}$ is guaranteed if $\hat{a}_{m,1}\le\bar{a}$.
This leads to the following stopping rule for the random walk algorithm.

\noindent
{\it {\bf Stopping rule 1.} Compute
$x_0=\arg\min\{x\in \mathbb{N}: (1-e^{-x})^k\ge 1-\bar\alpha/2.\}$
Stop at \[m_1=\arg\min\{m: X_{j_k}=x_0\}.\]}

We have observed in our numerical experiments that we obtain the best trade off between the number of steps
of the random walk and the accuracy if we take $\alpha$ around the average degree and the sampling probability $q$
around 0.5. Specifically, if we take $\bar{a}/2=0.15$ ($x_0=4$) in Stopping rule 1 for top 10 list,
we obtain 87\% accuracy for an average of 47\,000 random walk steps for the PA network;
92\% accuracy for an average of 174\,468 random walk steps for the DBLP network;
and 94\% accuracy for an average of 247\,166 random walk steps for the UK network.
We have averaged over 1000 experiments to obtain tight confidence intervals.

%Next, let us consider the random walk based sampling. Suppose we start the random walk
%with node $i$ and it is governed by the transition probability matrix $P$. Now define
%$$
%H_{ij} = \left\{ \begin{array}{ll}
%1, & \mbox{node $j$ has been observed at least once, if we start from node $i$,}\\
%0, & \mbox{node $j$ has not been observed, if we start from node $j$.}
%\end{array} \right.
%$$
%Let $\tilde P_{-j}$ be a substochastic matrix formed from $P$ by removing the $j$-th
%row and the $j$-th column. It represents an absorbing Markov chain with state $j$ as
%an absorption state. Then, the probability of not hitting state $j$ in $m$ steps is
%given by $e_i \tilde P_{-j}^m \ones$ and hence we have
%\begin{equation} \label{eq:numcorelrw}
%E[\sum_{j=1}^k H_{ij}] = \sum_{j=1}^k E[H_{ij}] = \sum_{j=1}^k (1-e_i \tilde P_{-j}^m \ones).
%\end{equation}
%There is a number of interesting questions: Are the eigenvalues of $\tilde P_{-j}$ well separated
%from one? What if we start from the stationary distribution or from the uniform distribution?

\section{Relaxation of top $k$ lists}

In the stopping criteria of the previous section we have strived to detect all nodes in
the top $k$ list. This costs us a lot of steps of the random walk.  We can significantly
gain in performance by relaxing this strict requirement.
For instance, we could just ask for list of $k$ nodes that contains 80\% of top $k$ nodes \cite{ALNSS11}.
This way we can take an advantage of a generic 80/20 rule that 80\% of result
can be achieved with 20\% of effort.

\begin{figure}[ht]
  \begin{center}
  \subfigure[$\alpha = 0$]{
    \includegraphics[scale=0.35]{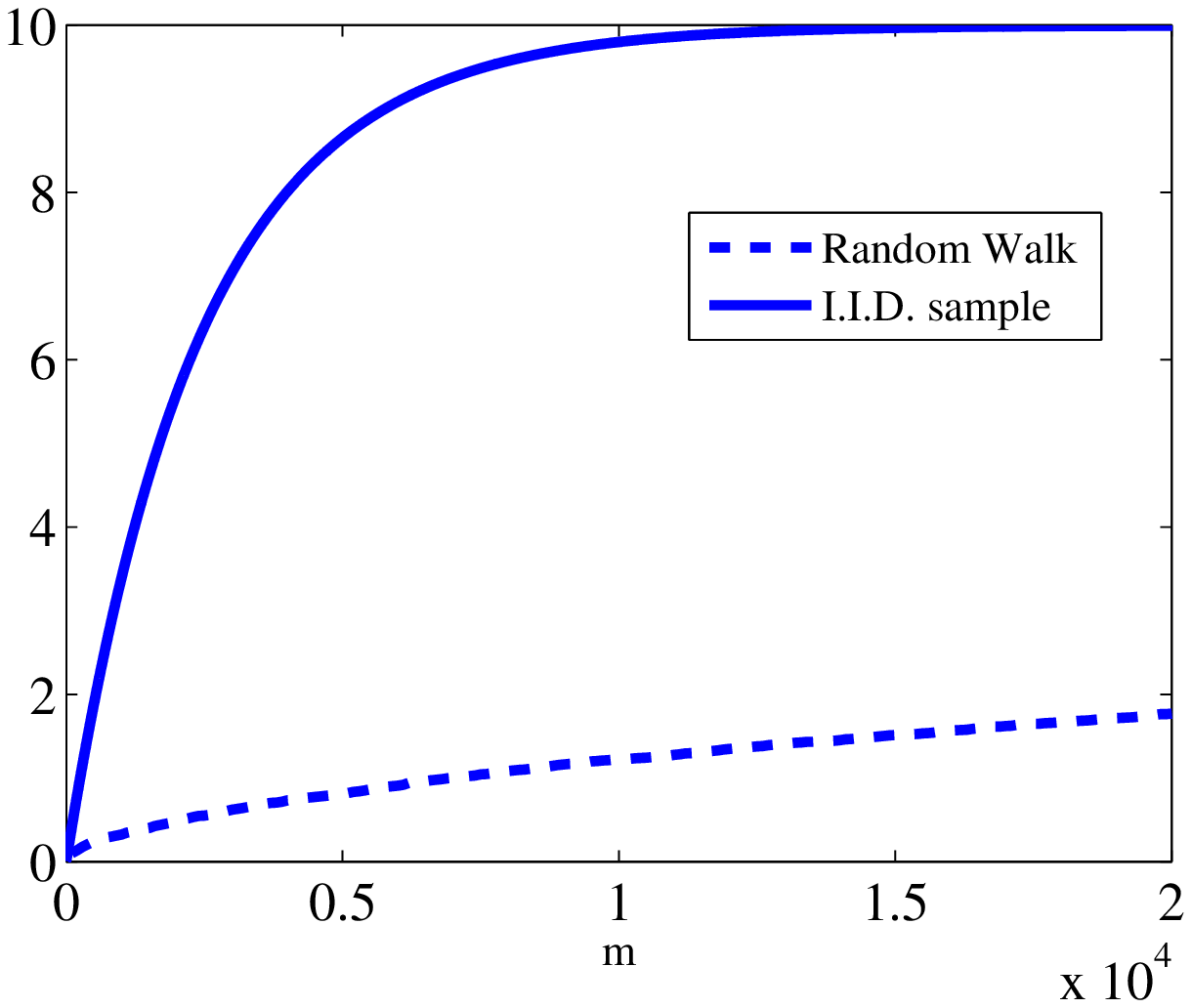}
  }
   \subfigure[$\alpha = 2$]{
   \includegraphics[scale=0.25]{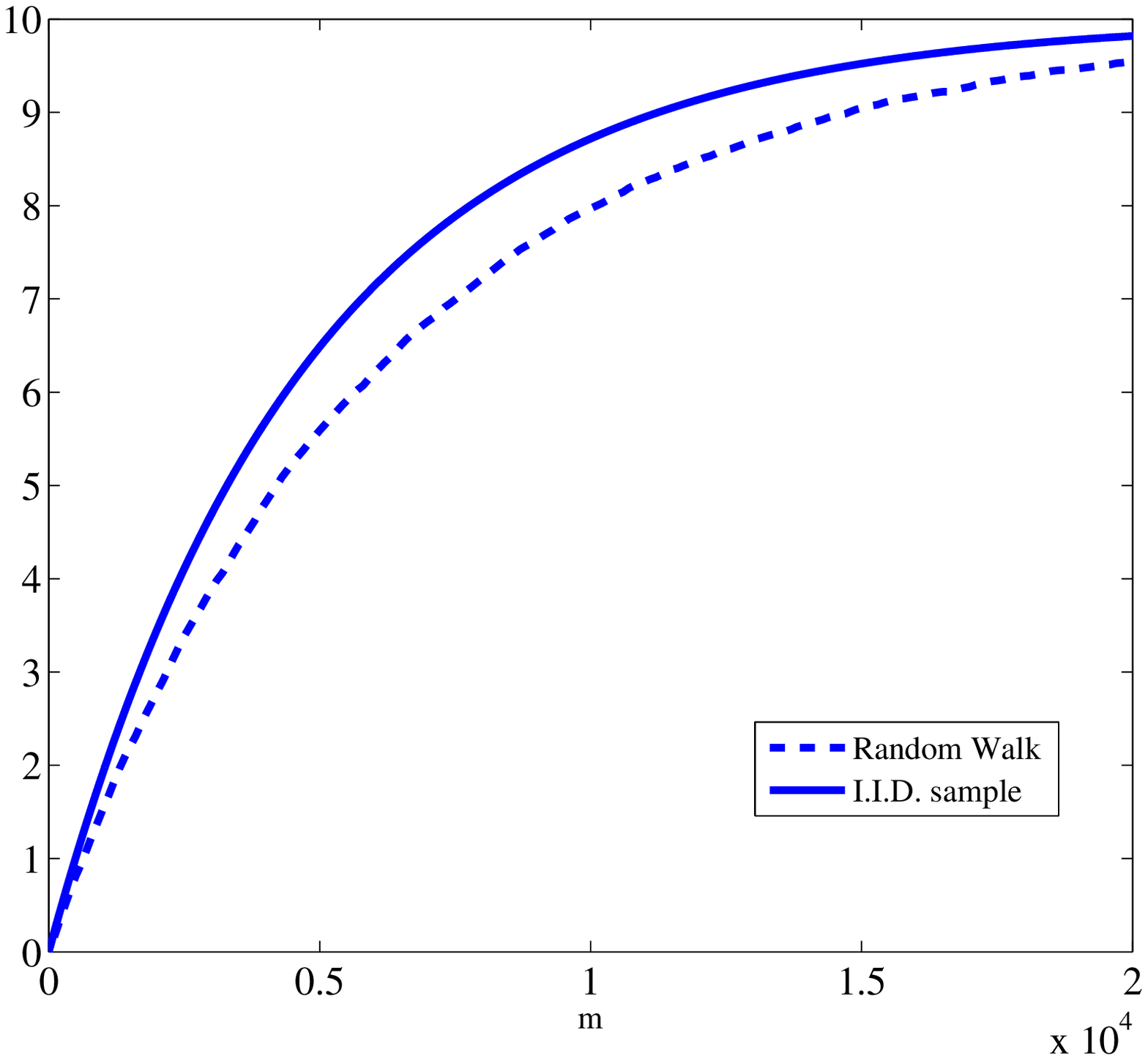}
   }
  \caption{Average number of correctly detected elements in top-10 for PA.\label{fig:NumElPA}}
  \end{center}
\end{figure}

\begin{figure}[ht]
  \begin{center}
  \subfigure[$\alpha = 0.001$]{
    \includegraphics[scale=0.27]{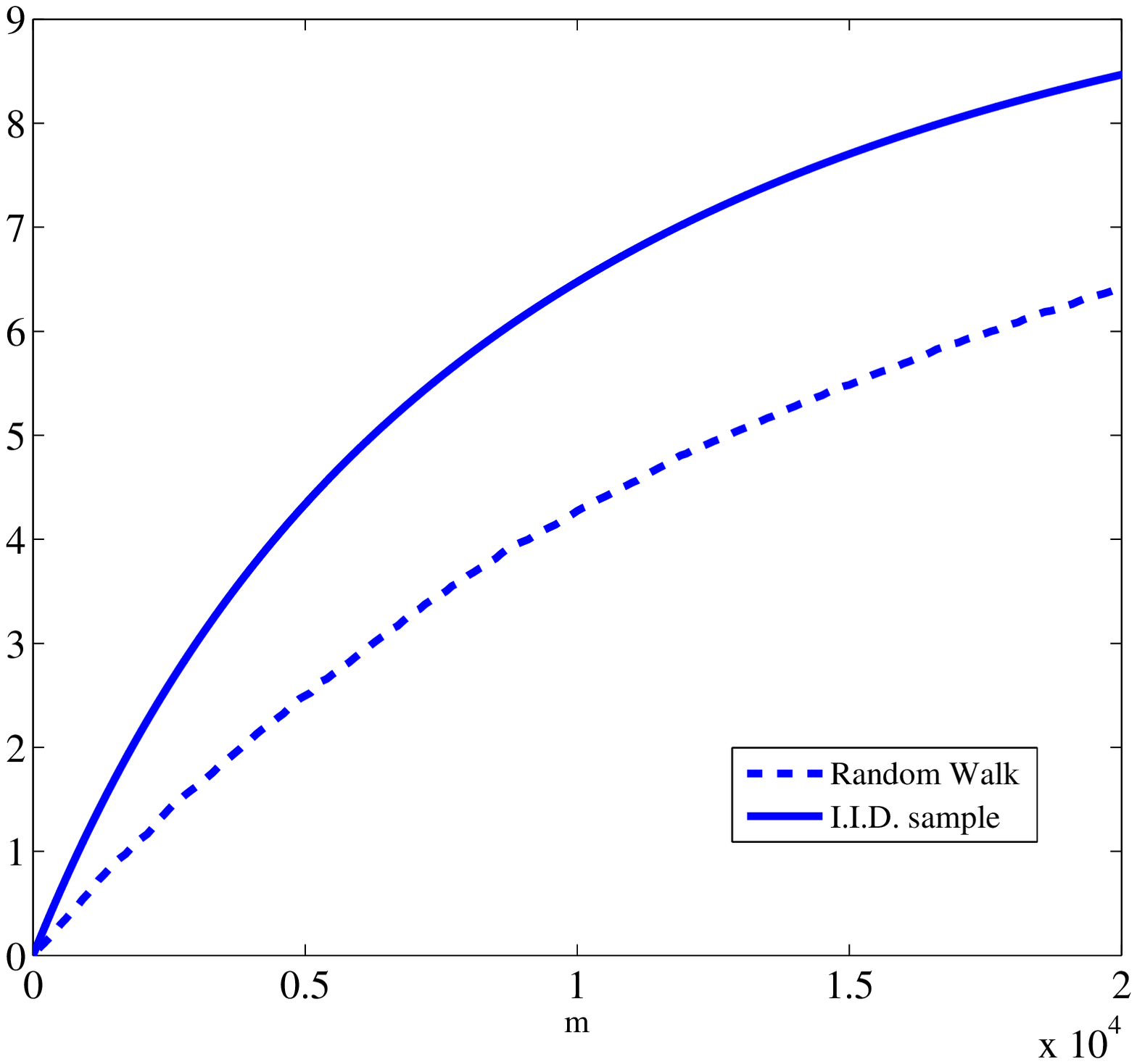}
  }
   \subfigure[$\alpha = 28.6$]{
   \includegraphics[scale=0.27]{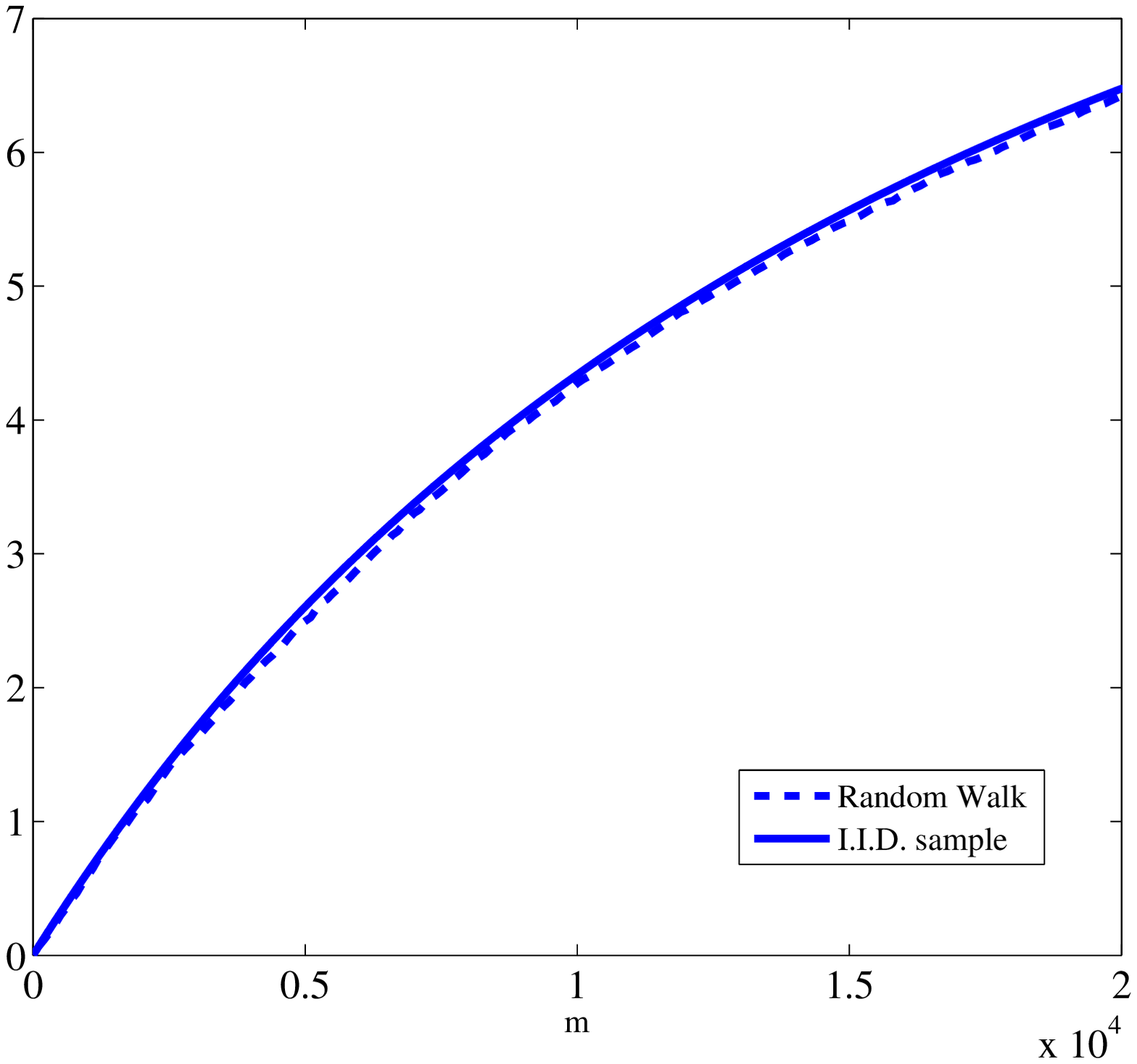}
   }
  \caption{Average number of correctly detected elements in top-10 for UK.\label{fig:NumElUK}}
  \end{center}
\end{figure}

Let us calculate the expected number of top $k$ elements observed
in the candidate list up to trial $m$. Define by $X_j$ the number of times
we have observed node $j$ after $m$ trials and
$$
H_j =
\left\{ \begin{array}{ll}
1, & \mbox{node $j$ has been observed at least once,}\\
0, & \mbox{node $j$ has not been observed.}
\end{array} \right.
$$
Assuming we sample in i.i.d. fashion from the distribution (\ref{eq:modstdistr}), we can write
\begin{equation} \label{eq:numcorel}
E[\sum_{j=1}^k H_j] = \sum_{j=1}^k E[H_j] = \sum_{j=1}^k P[X_j \ge 1] =
\sum_{j=1}^k (1-P[X_j = 0]) = \sum_{j=1}^k (1-(1-\pi_j)^m).
\end{equation}
In Figure~\ref{fig:NumElPA} we plot $E[\sum_{j=1}^k H_j]$ (the curve ``I.I.D. sample'') as a function of $m$
for $k=10$ for the PA network with $\alpha=0$ and $\alpha=2$.
In Figure~\ref{fig:NumElUK} we plot $E[\sum_{j=1}^k H_j]$ as a function of $m$ for $k=10$ for the UK network
with $\alpha=0.001$ and $\alpha=28.6$. The results for the UK and DBLP networks are similar in spirit.

Here again we can use the Poisson approximation
$$
E[\sum_{j=1}^k H_j] \approx \sum_{j=1}^k (1-e^{-m\pi_j}).
$$
In fact, the Poisson approximation is so good that if we plot it on Figures~\ref{fig:NumElPA}~and~\ref{fig:NumElUK},
it nearly covers exactly the curves labeled ``I.I.D. sample'', which correspond to the exact formula (\ref{eq:numcorel}).
Similarly to the previous section, we can propose stopping criteria based on the Poisson approximation. Denote
$$
b_{m} = \sum_{i=1}^k (1-e^{-X_{j_i}}).
$$

\noindent
{\it {\bf Stopping rule 2.} Stop at $m=m_2$, where
\[m_2=\arg\min\{m: b_{m}\ge \bar{b}\}.\]
}
\smallskip

Now if we take $\bar{b}=7$ in Stopping rule 3 for top-10 list,
we obtain on average 8.89 correct elements for an average of 16\,725 random walk steps for the PA network;
we obtain on average 9.28 correct elements for an average of 66\,860 random walk steps for the DBLP network; and
we obtain on average 9.22 correct elements for an average of 65\,802 random walk steps for the UK network.
(We have averaged over 1000 experiments for each network.)
This makes for the UK network the gain of more than two orders of magnitude in computational complexity
with respect to the deterministic algorithm.

\section{Conclusions and future research}

We have proposed the random walk method with the candidate list for quick detection
of largest degree nodes. We have also supplied stopping criteria which do not require
knowledge of the graph structure. In the case of large networks, our algorithm finds
top $k$ list of largest degree nodes with few mistakes with the running time orders
of magnitude faster than the deterministic sorting algorithm. In future research we
plan to obtain estimates for the required number of steps for various types of complex
networks.

\newpage
\tableofcontents

\end{document}